\documentclass[
    ,final            
  ]
  {aipproc}

\layoutstyle{6x9}

\newcommand{\beq}{\begin{equation}}
\newcommand{\eeq}{\end{equation}}
\newcommand{\bea}{\begin{eqnarray}}
\newcommand{\eea}{\end{eqnarray}}

\begin{document}

\title{Inclusive production of a pair of identified, rapidity-separated 
hadrons in proton collisions}


\author{D.Yu. Ivanov}{
  address={Sobolev Institute of Mathematics and
Novosibirsk State University, 630090 Novosibirsk, Russia}
}

\author{A. Papa}{
  address={Dipartimento di Fisica, Universit\`a della Calabria, 
and Istituto Nazionale di Fisica Nucleare, Gruppo collegato di
Cosenza, I-87036 Arcavacata di Rende, Cosenza, Italy}
}


\begin{abstract}
We consider the inclusive process where a pair of identified hadrons having 
large transverse momenta is produced in high-energy proton-proton collisions. 
We concentrate on the kinematics where the two identified hadrons in the final 
state are separated by a large interval of rapidity. In this case the cross 
section receives large higher-order corrections, which can be resummed in the 
BFKL approach. We provide a theoretical input for the resummation of such 
contributions with next-to-leading logarithmic accuracy. This process has much 
in common with the widely discussed Mueller-Navelet jets production and can be 
also used to access the BFKL dynamics at proton colliders.
\end{abstract}

\maketitle


\section{Introduction}

We consider  the semi-inclusive process 
\begin{equation}
{\rm proton}(p_1)   \, + \, {\rm proton}(p_2) \, \to \,
{\rm hadron}_1(k_1) \, + \, {\rm hadron}_2(k_2) \, + \, X \;.
\label{process}
\end{equation}
in the kinematics where the identified hadrons in the final state, $h_1$,
$h_2$, have large transverse momenta, $\vec k_1^{\:2}\sim \vec k_2^{\:2} \gg 
\Lambda_{\rm QCD}^2$, and are separated by a large interval of
rapidity $\Delta y\gg 1$ at $s=2 p_1\cdot p_2 \gg \vec k_{1,2}^{\:2}$. This 
process has much in common with the widely discussed Mueller-Navelet jet 
production~\cite{Mueller:1986ey} and can be also used to access the BFKL
dynamics~\cite{BFKL} at proton colliders.

In QCD collinear factorization the cross section of the process reads
\beq
\frac{d\sigma}{dy_1dy_2d^2 k_1 d^2 k_2} =\sum_{i,j=q,g}\int\limits^1_0
\int\limits^1_0 dx_1dx_2 f_i(x_1,\mu_F) f_j(x_2,\mu_F)
\frac{d\hat \sigma(x_1 x_2 s,\mu_F)}{dy_1dy_2d^2 k_1d^2 k_2}\;,
\label{ff}
\eeq
where the $i,j$-indices specify parton types, $i,j=q,\bar q, g$, $f_i(x,\mu_F)$
denotes the initial protons parton density functions (PDFs), the longitudinal
fractions of the partons involved in the hard subprocess are $x_{1,2}$,
$\mu_F$ is the factorization scale and $d\hat \sigma(x_1 x_2 s,\mu_F)$ is the
partonic cross section for the production of identified hadrons. The
latter is expressed in terms of parton fragmentation functions (FFs). The FF 
describes the probability of the inclusive fragmentation of the parton $i$ 
into a hadron $h$. The cross section is generically expressed as follows
\beq
d\sigma_i =C^h_i(z) dz\to d\sigma^h =d\alpha_h\int\limits^1_{\alpha_h}
\frac{dz}{z} D^h_i\left(\frac{\alpha_h}{z},\mu_F\right) C^h_i(z,\mu_F) \ ,
\label{fragm-r}
\eeq
where $C^h_i$ is the cross section (calculable in QCD perturbation theory) for
the production of a parton with momentum fraction $z$; the non-perturbative,
large-distance part of the transition to a hadron with momentum fraction
$\alpha_h$ is described in terms of the fragmentation function $D^h_i$;
$\mu_F$ stands here for the factorization scale.

In the considered Regge limit the partonic cross section can be calculated in  
the BFKL approach. Generically in BFKL the total cross section
$A + B \to X$ can be written as 
\begin{equation}
\sigma_{AB}=\frac{1}{(2\pi)^{2}}\!\int\!\frac{d^2 q_1}{\vec
q_1^{\,\, 2}}\Phi_A(\vec q_1,s_0)\!\int\!
\frac{d^2 q_2}{\vec q_2^{\,\,2}} \Phi_B(-\vec q_2,s_0)
\!\int\limits^{\delta +i\infty}_{\delta
-i\infty}\!\frac{d\omega}{2\pi i}\left(\frac{s}{s_0}\right)^\omega
G_\omega (\vec q_1, \vec q_2)\,.
\label{bfkl}
\end{equation}
This factorization is valid both in the leading logarithmic approximation
(LLA), which means resummation of all terms $(\alpha_s\ln s)^n$, and in the
next-to-LLA (NLA), which means resummation of all terms
$\alpha_s(\alpha_s\ln s)^n$. The Green's function $G_\omega$ is
process-independent and is determined through the BFKL equation
whose kernel is known in the NLA~\cite{FL+CC98}. As for the process-dependent 
impact factors (IFs) $\Phi_{A,B}$, only very few have been calculated in the 
NLA. Below we discuss our recent calculation~\cite{IP} of the
identified-hadron impact factor which is necessary for the NLA analysis of 
the process~(\ref{process}). 

We calculate the projection of the impact factor on the eigenfunctions of LO 
BFKL kernel, i.e. the impact factor in the so called $(\nu,n)$-representation,
\beq
\Phi(\nu,n)=\int d^2 q \,\frac{\Phi(\vec q\,)}{\vec q^{\,\, 2}}\frac{1}{\pi
\sqrt{2}}\left(\vec q^{\,\, 2}\right)^{i\nu-\frac{1}{2}} e^{i n \phi} \;.
\label{nu_rep}
\eeq
Here $\phi$ is the azimuthal angle of the vector $\vec q$ counted from
some fixed direction in the transverse space. This 
$(\nu,n)$-representation is the most convenient one in view of the
numerical determination of the cross section of process~(\ref{process}).

\section{The identified hadron impact factor}

The starting point for our NLA calculation is provided by the IFs for 
colliding partons~\cite{FFKP99+Cia}. In LLA quark and gluon IFs have the 
form ($N$ is the number of QCD colors)
\begin{equation}
\Phi_q=g^2\frac{\sqrt{N^2-1}}{2N}\;,\;\;\;\;\;\Phi_g=\frac{C_A}{C_F}\Phi_q\;,
\;\;\;\;\;\;\;\;\;\;C_A=N\;, \;\;\; C_F=\frac{N^2-1}{2N}\;.
\end{equation}
For the identified-hadron LLA IF in the $(\nu,n)$-representation we obtain 
the result
$$
\frac{\pi\sqrt{2}\, \vec k^{\, 2}}{{\cal C}}\frac{d\Phi^h(\nu,n)}
{d\alpha_h d^2 k}\!=\!\!\int\limits^1_{\alpha_h}\frac{dx}{x}
\!\left( \frac{C_A}{C_F}f_g(x) D_g^h\left(\frac{\alpha_h}{x}\right)
\!+\!\!\!\!\sum_{a=q,\bar q}f_a(x) D_a^h\left(\frac{\alpha_h}{x}\right)\right)
\!\left(\vec k^{\,\,2} \right)^{\gamma-{n \over 2}}
\left(\vec k \cdot \vec l \,\, \right)^n ,
$$
where $\gamma=i\nu-\frac{1}{2}, \alpha_s=g^2/(4\pi), {\cal C}=2\pi\alpha_s
\sqrt{\frac{2\, C_F}{C_A}}$, while $\vec k$ is the transverse momentum of the 
parton (quark or gluon) which fragments to the hadron $h$.

Note that for the LLA IF, there can be only a one-particle intermediate state, 
whereas for the NLA IF, we have to consider virtual corrections to the 
one-particle intermediate state, and also real particle production, with a 
two-particle intermediate state.
The intermediate steps of our calculation contain ultraviolet and infrared soft
and collinear divergences, which are regularized in the dimensional 
regularization.  
We have explicitly verified that in the final result for the IF soft and 
virtual infrared divergences cancel each other, whereas the infrared collinear 
ones are compensated by the PDFs' and FFs' renormalization counterterms, 
see~\cite{IP} for the details. The remaining ultraviolet divergences
are taken care of by the renormalization of the QCD coupling, which introduces 
the dependence on the renormalization scale $\mu_R$.

Another singularity which is present both for the partonic subprocesses 
initiated by the quark and the gluon PDFs is associated with the real particle 
production when an extra gluon is present in the intermediate state. It is 
related to the kinematic region when this extra gluon is emitted in
the central rapidity region, away from the fragmentation region of the initial 
quark/gluon. This contribution of the central region must be subtracted from 
the impact factor, since it is to be assigned in the BFKL approach to the 
Green's function, see~\cite{FF98} for the description of this subtraction in 
details. 

Finally, we arrive at our NLA result for the identified-hadron IF
\[
\vec k^{\, 2}_h \,\,\frac{d\Phi^h(\nu,n)}
{d\alpha_h d^2 k_h}=2\,\alpha_s(\mu_R)\sqrt{\frac{C_F}{C_A}}
\left(\vec k_h^{\,\,2}
\right)^{\gamma-{n \over 2}}\!\left(\vec k_h \cdot \vec l \, \right)^n
\!\left\{
\int\limits^1_{\alpha_h}\frac{dx}{x} \left(\frac{x}{\alpha_h}\right)^{2\gamma}
\!\left[ \frac{C_A}{C_F}f_g(x)
D_g^h\left(\frac{\alpha_h}{x}\right) \right.\right.
\]
\begin{equation}
\left. +\sum_{a=q,\bar q}f_a(x)
D_a^h\left(\frac{\alpha_h}{x}\right)\right]
+ \frac{\alpha_s\left(\mu_R\right)}{2\pi}\int\limits^1_{\alpha_h} \frac{dx}{x}
\int\limits^1_{\frac{\alpha_h}{x}}
\frac{d\zeta}{\zeta}\left(\frac{x\, \zeta}{\alpha_h}\right)^{2\gamma}
\end{equation}
\[
\times
\left[\frac{C_A}{C_F}f_g(x)
D_g^h\left(\frac{\alpha_h}{x\zeta}\right)
C_{gg}\left(x,\zeta\right)+\sum_{a=q,\bar q}f_a(x)
D_a^h\left(\frac{\alpha_h}{x\zeta}\right)
C_{qq}\left(x,\zeta\right) \right.
\]
\[
\left.\left.
+\sum_{a=q,\bar q}f_a(x)
D_{g}^h\left(\frac{\alpha_h}{x\zeta}\right)
C_{qg}\left(x,\zeta\right)
+\frac{C_A}{C_F} f_{g}(x)\sum_{a=q,\bar q}
D_a^h\left(\frac{\alpha_h}{x\zeta}\right)
C_{gq}\left(x,\zeta\right) \right]
\right\}\;.
\]
The explicit form of the coefficient functions $C_{ij}$ is given in our 
paper~\cite{IP}.

\section{Summary}

We have calculated the NLO impact factor for the forward production of an 
identified hadron from an incoming quark or gluon, emitted by a proton. 
This is a necessary ingredient for the calculation of the
hard inclusive production of a pair of rapidity-separated identified hadrons
in proton collisions~(\ref{process}). This process, similarly to the
production of Mueller-Navelet jets, can be studied at the LHC hadron collider.

Another natural application of the obtained identified hadron production IF
could be the NLA BFKL description of inclusive forward hadron production
process in DIS,
\beq
e(p_1)+p(p_2)\to h(k)+X \ ,
\label{process1}
\eeq
where in the low-$x$ event the hadron $h(k)$ with high transverse momentum is
detected in the fragmentation region of incoming proton $p(p_2)$. Data
for such reaction in the case of forward $\pi^0$-production were published by
the H1 collaboration at HERA~\cite{Aktas:2004rb}.

In our approach the energy scale $s_0$ (which enters~(\ref{bfkl})) is an 
arbitrary parameter, that need not be fixed at any definite scale. The 
dependence on $s_0$ will disappear in the next-to-leading logarithmic 
approximation in any physical cross section in which the identified-hadron 
production vertices are used. Indeed, due to discussed above subtraction 
of the central rapidity gluon emission, our result for the NLO vertex contains 
contributions $\sim \ln(s_0)$, 
$$
C_{\{gg\}, \{qq\}}\left(x,\zeta\right) =\delta(1-\zeta) C_A \ln\left(\frac{s_0 \, \alpha_h^2}{\vec k^2_h \,
x^2 }\right)\chi(n,\nu) + \dots  \ .
$$
Note that these terms are proportional to the LO quark and gluon vertices 
multiplied by the BFKL kernel eigenvalue $\chi(n,\nu)$. This fact guarantees 
the independence of the identified-hadrons~(\ref{process}) or 
single-hadron~({\ref{process1}}) cross section on $s_0$ within the 
next-to-leading logarithmic approximation.
However, the dependence on this energy scale will survive in terms beyond
this approximation and will provide a parameter to be optimized with the
method adopted in Refs.~\cite{mesons,Caporale:2012ih}.


\begin{theacknowledgments}
{\bf Acknowledgments.} D.I. thanks the Dipartimento di Fisica
dell'Universit\`a della Calabria and the Istituto Nazionale di Fisica
Nucleare (INFN), Gruppo collegato di Cosenza, for the warm hospitality
and financial support. This work was also supported in part by the
grants RFBR-11-02-00242 and NSh-3810.2010.2.
\end{theacknowledgments}

\end{document}